\documentclass[twoside,12pt]{article}

\usepackage{cite}
\usepackage{epsfig}
\usepackage{amsmath}
\usepackage{amssymb}

\newcommand{\be}{\begin{equation}}
\newcommand{\ee}{\end{equation}}
\newcommand{\bea}{\begin{eqnarray}}
\newcommand{\eea}{\end{eqnarray}}

\begin{document}

\title{ \vspace{1cm} 
		The Proton Radius Puzzle}
\author{Carl E. Carlson\\
\\
Physics Department, College of William and Mary, Williamsburg, VA 23187, USA}
\maketitle
\begin{abstract} 
The proton size, specifically its charge radius, was thought known to about 1\% accuracy.  Now a new method probing the proton with muons instead of electrons finds a radius about 4\% smaller, and to boot gives an uncertainty limit of about 0.1\%.  We review the different measurements, some of the calculations that underlie them,  some of the suggestions that have been made to resolve the conflict, and give a brief overview new related experimental initiatives.  At present, however, the resolution to the problem remains unknown.
\end{abstract}

\tableofcontents

\section{Introduction}
The proton radius conflict is a prime problem in proton structure physics.  To state the problem is simple:  we measure the proton radius using electrons, and we measure the proton radius using muons, and we get incompatibly different answers.

Measurements made using electrons come either from electron-proton scattering or from atomic spectroscopy, where there are small but measurable shifts in the hydrogen spectrum due to the proton size.  The proton charge radius obtained from different electron measurements substantially agree, and gave a result with an uncertainly of order 0.6\%.

On the muonic side, there is only one set of measurements, but it is one with a very small uncertainty limit.  The announcement came in 2010~\cite{Pohl:2010zz}, with follow-up in 2013~\cite{Antognini:1900ns}, of the proton size measured from its effect on the muonic hydrogen spectrum, specifically the 2S-2P Lamb shift.  The muon, because of its mass, orbits closer to the proton than an electron, and it energy levels are more strongly affected by the proton size.  This allows a proton radius measurement with an error bar more than 10 times smaller than obtained from the collected electronic measurements, despite the limited lifetime of the muon.

The proton radius from the muonic Lamb shift is a startling 4\%, or 7 of the current electronic standard deviations, smaller than the previously accepted result.

The reasons for this are not yet clear.  We will make some comments here about what has been thought about, beginning with a brief description of the electron based experiments followed by some remarks about how the muonic Lamb measurements are carried out to give the charge radius.    We will then comment on some (but far from all) of the theory calculations needed to interpret the experiments.  Much work has been done checking the corrections involved in pulling out the charge radius from the muonic experiment,  but nothing big enough to affect the result that has been found.  Then we will comment at some length about suggestions that have been made to reconcile the two measurements, including speculation about anomalous QCD corrections, speculation about connections to physics beyond the standard model, and questions raised about the interpretation of the experiments using electrons.  The importance of the problem has led to a number of experiments to check aspects of the conflict, and we will catalog some of them in anticipation of their success.  We will offer some concluding remarks, but the main concluding remark, unfortunately or otherwise, is that the problem remains. 

Previous reviews are found in~\cite{Jentschura:2010ej,Pohl:2013yb}.


\section{The electronic measurements}


Measurements of the proton radius using electrons, either by electron-proton scattering or by careful measurements of atomic energy levels, did not originate the proton radius puzzle.  The various electron based measurements of the radius are  in agreement and it was not until the muon based measurement appeared that there was a problem.  However, we will start with a discussion of the electron based measurements in order to show our notation and delineate one side of the conundrum.

We start with electron scattering.  For understanding some definitions, it is even useful to start non-relativistically, where scattering electrons off pointlike protons has the Rutherford cross section
\begin{equation}
\left. \frac{d\sigma}{d\Omega} \right|_{\rm point}
= \left( \frac{\alpha}{4 E \sin^2(\theta/2)} \right)^2	\,,
\end{equation}
in the lab system, where $E$ is the energy of the incoming electron, $\theta$ is its scattering angle, $\alpha$ is the fine structure constant, $e^2/(4\pi)$.  The nonrelativistic cross section result off an extended target is modified, but just becomes
\begin{equation}
\frac{d\sigma}{d\Omega}
= \left. \frac{d\sigma}{d\Omega} \right|_{\rm point}	\times \Big( G(Q^2) \Big)^2	\,,
\end{equation}
where $Q$ is the momentum transfer, the difference between the incoming and outgoing projectile momenta, and $G(Q^2)$ is the ``form factor'', given nonrelativistically by
\begin{equation}
\label{eq:FF}
G(Q^2) \stackrel{NR}{=} \int d^3r \ 
e^{\,i \vec Q \cdot \vec r} \ 
| \psi (r) |^2	\,,
\end{equation}
where $\psi(r)$ is a wave function describing the density of the target.
It is easy to expand the form factor at low $Q^2$ and obtain,
\begin{equation}
G(Q^2) = 1 - \frac{1}{6} 
\langle r^2 \rangle Q^2 + \ldots ,
\end{equation}
where $\langle r^2 \rangle$ is the rms radius squared.

\begin{figure}[tb]
\begin{center}
\begin{minipage}[t]{11 cm}
\centerline{  \epsfig{file=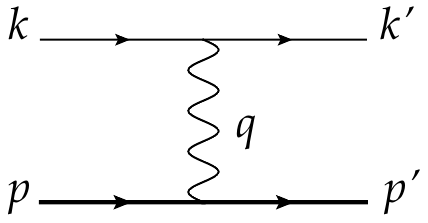,scale=1.0}  } 
\end{minipage}
\begin{minipage}[t]{16.5 cm}
\caption{One photon exchange.\label{fig:exchanges}}
\end{minipage}
\end{center}
\end{figure}

For a electron-proton scattering at the energies under consideration, relativity matters.
The lowest order calculation is one photon exchange and proceeds from the Feynman diagram in Fig.~\ref{fig:exchanges}.  The proton vertex involves the matrix element of the electromagnetic current between incoming and outgoing states of definite momentum,
\begin{equation}
\langle p' | J_\mu  | p \rangle
= \bar u(p') \left( \gamma_\mu F_D(Q^2)+ \frac{i}{2m_p} \sigma_{\mu\nu}\,
q^\nu F_P(Q^2) \right)  u(p)	\,,
\end{equation}
where $\sigma_{\mu\nu} = (i/2) [\gamma_\mu,\gamma_\nu]$, $q = p'-p$, $F_D$ and $F_P$ are the Dirac and Pauli form factors, spin degrees of freedom are tacit, and $Q^2 = - (p'-p)^2$.

The electric and magnetic form factors are linear combinations of $F_D$ and $F_P$,
\begin{align}
G_M &= F_D + F_P \,, \\
G_E &= F_D - \tau F_P  \,,
\end{align}
and the differential cross section to leading order in perturbation theory is given by
\begin{equation}
\frac{d\sigma}{d\Omega} = 
\left. \frac{d\sigma}{d\Omega} \right|_{NS}\times
\frac{1}{ (1+\tau)}
		\Big(  G_E^2(Q^2) +  
\frac{\tau}{ \epsilon}G_M^2(Q^2)  \Big)	\,.
\end{equation}
Here and above,
\begin{align}
\tau &= \frac{Q^2 }{ 4M^2}	\,,	\nonumber\\
\frac{1 }{ \epsilon} &= 1 + 2(1+\tau) \tan^2 \frac{\theta}{ 2}	\,,
\end{align}
and 
and $\left. {d\sigma / d\Omega} \right|_{NS}$ is the Mott or  ``no-structure'' cross section (specifically, the hypothetical cross section for scattering an unpolarized spin-1/2 electron from a pointlike spin-0 target),
\begin{equation}
\left. \frac{d\sigma }{ d\Omega} \right|_{NS}
		= \frac{ 4\alpha^2  \cos^2\frac{\theta}{ 2}  }{ Q^4 } \,
		\frac{E^{\prime 3} }{ E}  \ ,
\end{equation}
where $E$ and $E'$ are the incoming and outgoing electron laboratory energies.
Obtaining the form factors by measuring the differential cross section at a variety of $Q^2$ and scattering angles is termed a Rosenbluth separation.

Relativistically, Eq.~(\ref{eq:FF}) is an approximation valid if the target's Compton wavelength is small compared to its size.  For a proton the latter statement is marginal.  Instead we define the form factors from the matrix element given above, and for the electric and magnetic radii, we promote the former derived result to a definition,
\begin{equation}
R_E^2 \stackrel{def}{=} -6 \left. 
\frac{dG_E}{dQ^2} \right|_{Q^2=0}	\,,
\end{equation}
with a similar relation between $R_M$ and the magnetic form factor.

The most accurate experiments for obtaining the form factors at low and moderate $Q^2$ are done at Mainz and reported in~\cite{Bernauer:2010wm}.   They quote their results by fitting form factors using a variety of different fitting functions to their measured differential cross sections. Their fits lead to
\begin{equation}
R_E = 0.879 (8) {\rm\ fm}	\,,
\end{equation}
where, following others~\cite{Walcher:2012qp}, several uncertainties are combined into a single uncertainty limit.

The other way to obtain the proton radius using electrons is to measure the energy levels, the Lamb shift but not only the Lamb shift, in ordinary electronic hydrogen. The proton radius measurements made using ordinary hydrogen are quite remarkable because the proton radius effects are very small.   Energy level splittings can be measured so precisely, and the proton size independent energies can be calculated so reliably,  that proton size dependent terms can be isolated.  

\begin{figure}[tb]
\begin{center}
\begin{minipage}[t]{8 cm}
\epsfig{file=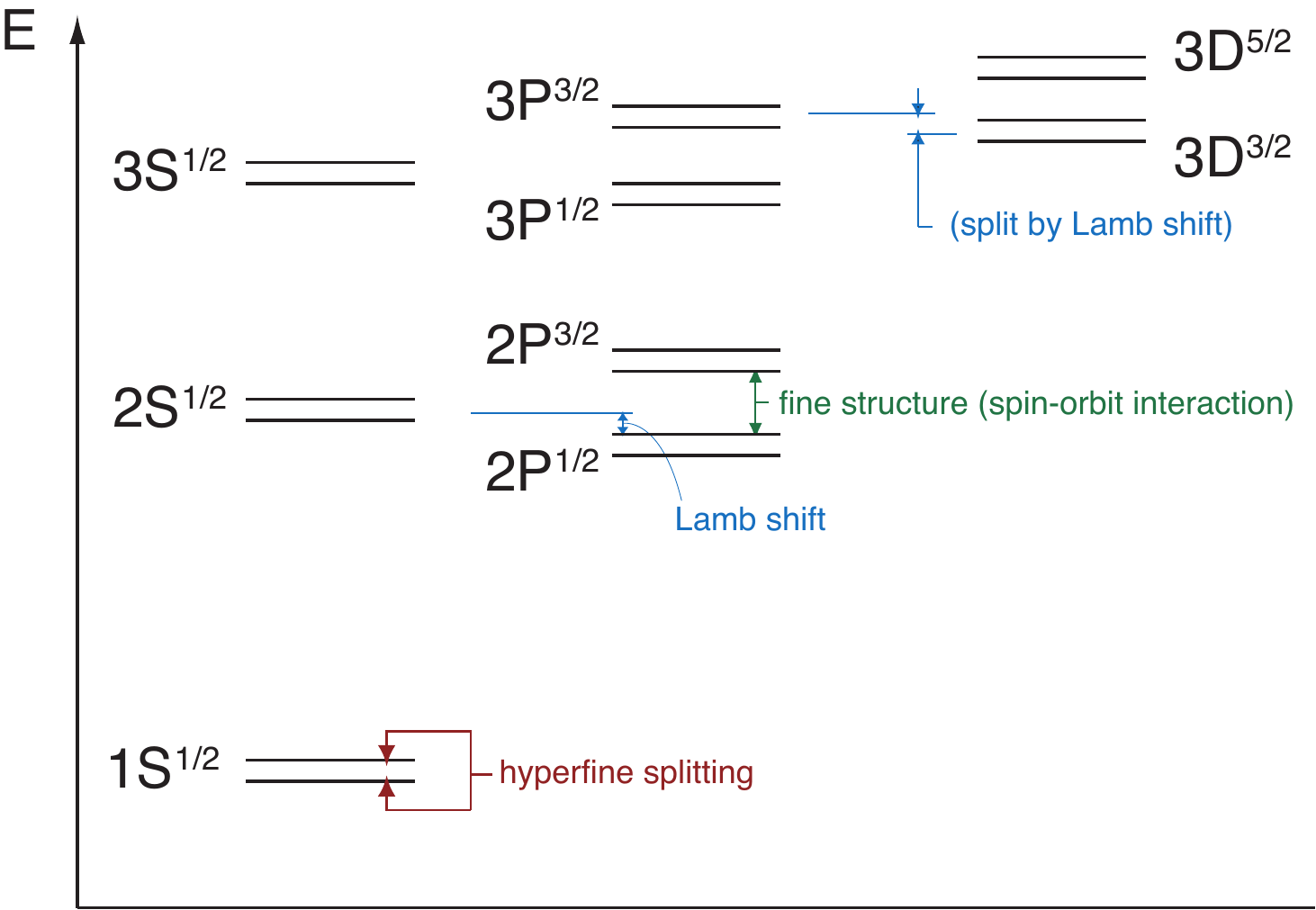,scale=0.55}
\end{minipage}
\begin{minipage}[t]{16.5 cm}
\caption{Spectrum of ordinary hydrogen (not to scale).\label{fig:H-spectrum}}
\end{minipage}
\end{center}
\end{figure}

As a reminder, the proton energy spectrum is illustrated in Fig.~\ref{fig:H-spectrum}.  The Figure is not to scale.  (That the Lamb shift---the splitting between the 2S$_{1/2}$ and 2P$_{1/2}$ levels---is about 10\% of the 2P fine structure splitting is the only relative splitting that is about right.)

The energy levels, at least the S-state levels, are affected by the finite proton size.  The size dependent energy shift was worked out non-relativistically in~\cite{Karplus:1952zza}, and is to leading order in perturbation theory (in modern notation)
\begin{equation}
\Delta E_1 = \frac{2 \pi \alpha}{3} | \phi^2(0) | R_E^2	\,,
\end{equation}
where $\phi(0)$ is the electron wave function at the origin in coordinate space, 
$\phi^2(0) = (m_r \alpha)^3/(\pi n^3)$, where $m_r$ is the reduced mass and $n$ is the principal quantum number.  Interestingly, the result depends only on the RMS proton radius squared and not on the detailed distribution of matter within the proton.
Relativistic considerations may be examined in~\cite{Eides:2000xc,Jentschura:2010ty}, and the result is that the formula is the same with the $R_E^2$ appearing being precisely the $R_E^2$ defined from the derivative of the electric form factor.

For the discussion, the measurements of the charge radius from atomic energy splittings can be divided into small splitting measurements, which are measurements of the corrections to the Lamb shift, all levels having the same principal quantum number, and big splitting measurements, which are corrections to splittings between levels with different principal quantum numbers.  Every splitting has a QED term, which is what the splitting would be if the proton were pointlike, and the QED term is written as some factor times the Rydberg constant, which is the leading order binding energy of the ground state hydrogen atom in a infinite proton mass limit (and in some units is $\alpha^2 m_e c^2/2$).  Also, every splitting has a proton radius term, which is noticeable only if a S-state is involved.  Generically,
\begin{equation}
\Delta E = a \cdot Ryd + b R_E^2	\,.
\end{equation}
The coefficients $a$ and $b$ are known;  for $a$, which must be known to high accuracy, this entails extensive calculation.

For the small splitting measurements, the Rydberg constant is accurately enough known from other sources to subtract the Rydberg constant term from $\Delta E$ and obtain the $R_E^2$ term.  For the big splitting measurements, the $R_E^2$ term is tiny compared to the main term and the Rydberg constant is not independently known well enough to accurately obtain the $R_E^2$ term from a single measurement.  The solution is to measure two splittings, and solve the elementary pair of simultaneous equations to obtain both the Rydberg constant and $R_E$.  To repeat: \textit{the high precision value of the Rydberg constant is obtained from the very same atomic energy level experiments that measure the proton radius}.  One gets the Rydberg constant to about 5 parts per trillion ($10^{12}$) and the proton radius to a bit better than 1\% when all averaging is done~\cite{Mohr:2012tt}.

\begin{figure}[tb]
\begin{center}
\begin{minipage}[t]{8 cm}
\centerline {  \epsfig{file=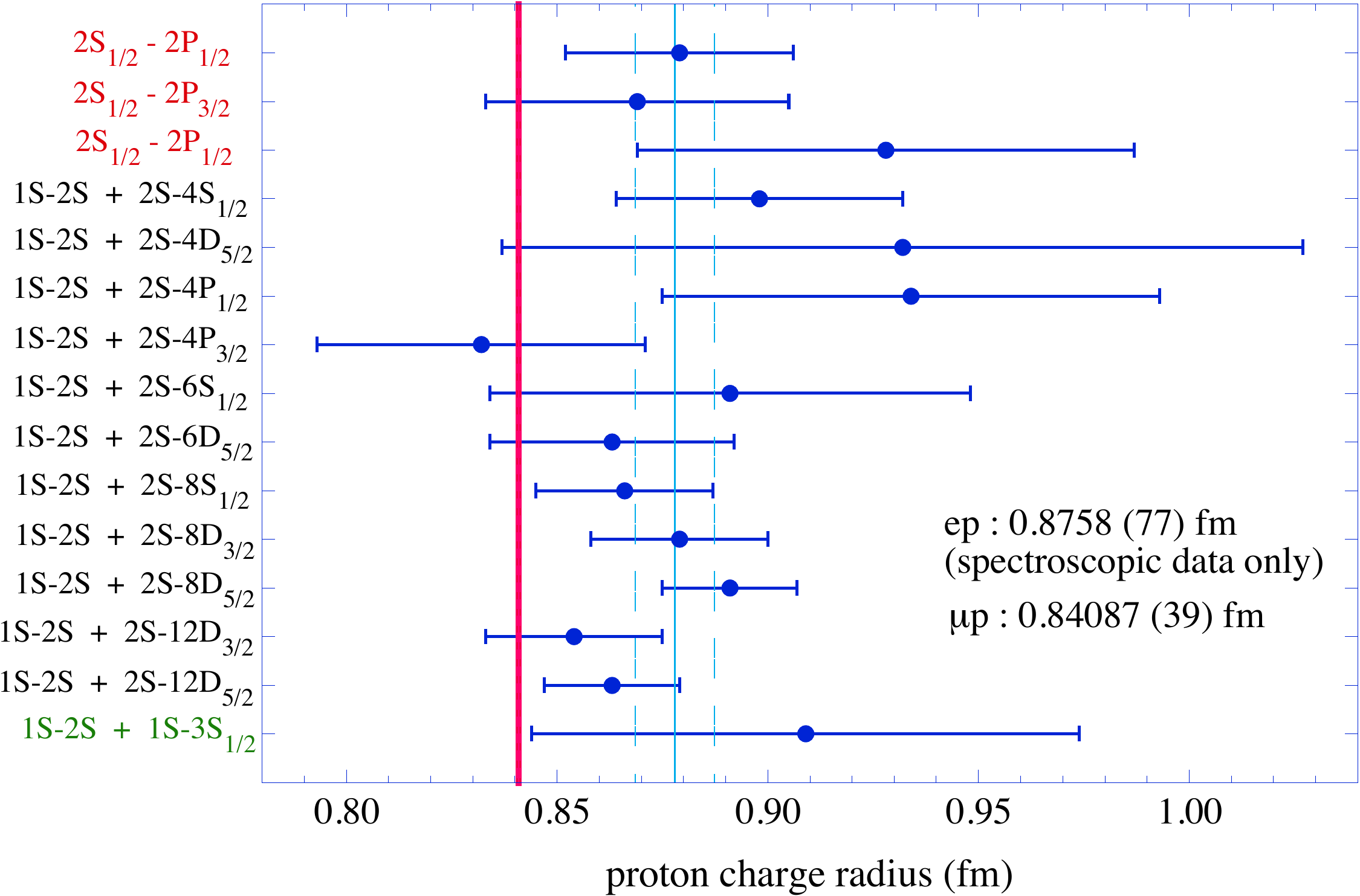,scale=0.5}  }
\end{minipage}
\begin{minipage}[t]{16.5 cm}
\caption{Proton radius from ordinary hydrogen level splitting measurements.\label{fig:Hradiusmeaurements}}
\end{minipage}
\end{center}
\end{figure}

A summary of the experimental hydrogen atom results in graphical form is shown in Fig.~\ref{fig:Hradiusmeaurements}.  The top three results are Lamb shift measurements.  The next 12 are from pairs of larger splittings.  The $1S$-$2S$ splitting is very accurately measured and is always used as one member of the pair;  after that there is wide selection of level splittings.  In general the spectral lines have a natural width because the states can decay, and to obtain the quoted accuracy, the centers of the lines must be located to a part in 100 to a part in 1000 of the natural width.  One notices immediately that none of the results is a 1\% measurement.  The sub-1\% uncertainty limit comes from averaging many experiments.  

The proton radius measurements using electrons are summarized (as part of a extensive summary of physical and technical data) by the Committee on Data for Science and Technology (CODATA), which has been updating results every 4 years.  When the first muonic results appeared, the 2006 CODATA results were available.  These included some electron scattering results with about 2\% uncertainty limits~\cite{Sick:2003gm}, but were dominated by the atomic measurements and reported $R_E = 0.8768(69)$ fm.  The up-to-date number is the CODATA 2010 value, which by the CODATA protocol is based on all data available through the end of 2010 and so was not available in 2010 itself.  The current CODATA average using only electronic spectroscopy data is $0.8758(77)$ fm~\cite{Mohr:2012tt}  .  With the Mainz electron scattering result included, CODATA finds the overall result~\cite{Mohr:2012tt},
\begin{equation}
R_E = 0.8775(51) \ {\rm fm}.
\end{equation}

The electron based proton radius results are mutually compatible.  Within the available accuracy, the electron based experiments by themselves create no problems.  However, there came the tempting prospect of increasing the accuracy by a factor of ten,  and pursuing that prospect led to a great surprise and the present day conundrum.

\section{The muonic Hydrogen experiment}

The muon is, as far as we know today, like an electron in all respects except in its mass or things directly related to the mass.  in particular, one can make a hydrogen atom using a negative muon in place of an electron.  The muon is about 200 times heavier than the electron, and in an atomic bound state it lies about 200 times closer to the proton than the electron, so proton size effects are greatly magnified.  

Despite the muon's limited 2.2 $\mu$s lifetime, it was anticipated that the larger impact of the proton size on the energy levels would allow a $0.1$\% measurement of the proton charge radius.   In 2010, the promise was more than realized, as a collaboration of experimenters announced their first result~\cite{Pohl:2010zz}, based on measuring one $2S$-$2P$ energy splitting.  The same collaboration has more recently published measurements of a second splitting, and slightly improved the analysis of the first, with the current result~\cite{Antognini:1900ns},
\begin{equation}
\label{eq:re}
R_E = 0.84087(39) {\rm\ fm}
\end{equation}
This is 4\% smaller than the CODATA radius, and is a 7$\sigma$ discrepancy. 

The splittings measured in the muonic experiment are splittings in the $2S$-$2P$ system, \textit{i.e.}, they are Lamb shift measurements, and so are the analogs of the ``small splitting'' measurements described in the electron section.  The experiment did not prove easy to make work, but once working, the description of it in outline is straightforward.  

\begin{figure}[tb]
\begin{center}
\begin{minipage}[t]{12 cm}
\epsfig{file=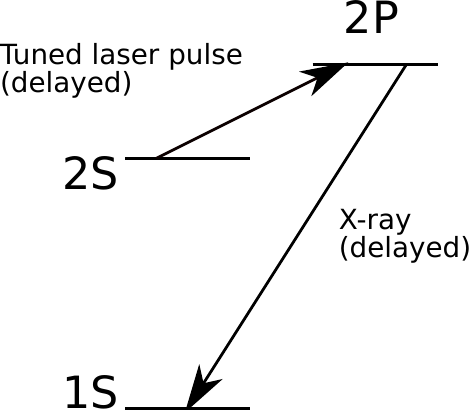,scale=1.1} \hfill
\epsfig{file=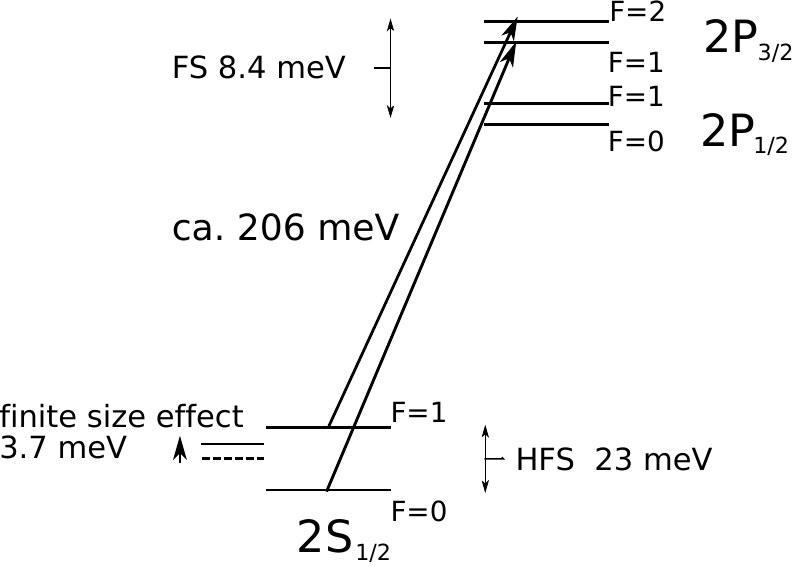,scale=0.78}
\end{minipage}
\begin{minipage}[t]{16.5 cm}
\caption{Energy levels in muonic hydrogen.  On the left, the 2S-2P transition is induced by a tuned laser, with success indicated by an X-ray from the 2P-1S transition.  On the right is a detail of the $n$=$2$ energy levels of muonic hydrogen.\label{fig:pstate}}
\end{minipage}
\end{center}
\end{figure}

The experiment is done at the Paul Scherrer institute in Switzerland by the CREMA (Charge Radius Experiment with Muonic Atoms) collaboration and begins with a beam of muons that are slowed and then sent into a hydrogen target, where they are finally stopped and captured on hydrogen atoms in a variety of orbits.  The muons cascade down very quickly, and nearly all the muons cascade down to the $1S$ state, with  few stuck in the $2S$ state, which is metastable.  Only about 1\% of the muons are in the $2S$ state, but these are the ones of interest.  After a short delay, to allow the aforementioned cascades to finish, the experimenters shine a tunable laser on the hydrogen, and if the frequency of the laser light is just right,  there will be a transition to a $2P$ state; see the left hand panel of Fig~\ref{fig:pstate}.  

The $2P$ state decays very quickly to the ground state, emitting an X-ray, and one has an X-ray detector to signal success in tuning the laser correctly.  Then the laser frequency gives the desired energy splitting.  The value of $\hbar$, incidentally, is only known to a bit better than 7 figures~\cite{Mohr:2012tt};  however, 6 figures suffice for the case at hand.

The two lines that are measured are indicated in the right hand panel of Fig.~\ref{fig:pstate}, where the $2S$ and $2P$ states of muonic hydrogen are shown in more detail.  (The subscripts on $2S$ and $2P$ indicate the summed angular momentum of the orbital motion and electron spin, and $F$ indicates the total angular momentum, orbital motion plus electron spin plus proton spin.)  The $P$-states are not measureably affected by proton size effects, but the $S$-states are, and the proton size affects the hyperfine splitting as well as to the overall $S$-state energy level.  The strict definition of the Lamb shift is the $2S_{1/2}$-$2P_{1/2}$ splitting with hyperfine splitting effects removed.  (The last is easy to do, since the spin-spin interaction moves the $F=1$ states up by 1/4 of the overall splitting and the $F=0$ states down by the complementary amount; for more detail see~\cite{Jentschura:1997hy}.)

The two splittings measured are the original  $2S_{1/2}^{F=1}$ to $2P_{3/2}^{F=2}$ and the more recent $2S_{1/2}^{F=0}$ to $2P_{3/2}^{F=1}$, with splittings in energy units denoted $h \nu_t$ and $h \nu_s$, respectively.   With two splittings one can obtain both the Lamb shift and the hyperfine splitting.  The former is the main goal of the experiment, but the hyperfine splitting also has proton structure effects, this time dependent on the Zemach radius~\cite{Zemach:1956zz}, to be discussed below, and the Zemach radius can be determined from this experiment, albeit not with the same precision as the charge radius.

The Lamb shift and hyperfine splitting are
\begin{align}
E_{\rm L} &= \frac{1}{4} h \nu_s + \frac{3}{4} h \nu_t - 8.8123(2) \rm{\, meV},	\nonumber\\
E_{\rm HFS} &= h \nu_s - h \nu_t +3.2480(2) \rm{\, meV},
\label{eq:hfs}
\end{align}
where the numerical terms follow from reliable (proton structure independent) calculations of the P-state energy splittings.

\begin{figure}[tb]
\begin{center}
\begin{minipage}[t]{11 cm}
\centerline{ \epsfig{file=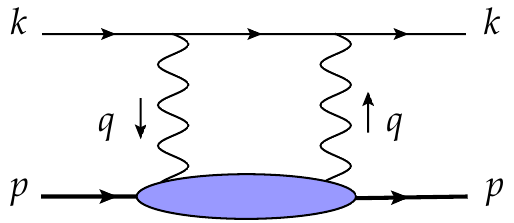,scale=1.0} }
\end{minipage}
\begin{minipage}[t]{16.5 cm}
\caption{Two-photon exchange correction.\label{fig:tpe}}
\end{minipage}
\end{center}
\end{figure}

Experimentally, 
\begin{align}
E_{\rm L}^{\rm exp} = 202.3706(23) {\rm\, meV}.
\end{align}
From theory,
\begin{align}
E_{\rm L}^{\rm th} = 206.0336(15) - 5.2275(10) R_E^2 + E_{\rm TPE}	\,,
\end{align}
where the energies are in meV and $R_E$ is in fm.  The first numerical term includes a number of corrections that are independent of the proton charge radius, and the last term is the two-photon correction, Fig.~\ref{fig:tpe}.  The latter will be discussed in its own section below.  For now we will mention that if the blob between the two photon attachments on the proton side is restricted to being a proton itself, and if the calculation is done non-relativistically, the result is proportional to the average separation-cubed of the proton charge, denoted $R^3_{(2)}$.  One can find it so written in older writing (\textit{e.g.,}~\cite{Friar:1978wv,Pohl:2010zz}).  Today one uses the value
\begin{equation}
E_{\rm TPE} = 0.0332(20) \rm{\, meV}
\label{eq:tpe}
\end{equation}
based largely on calculations in~\cite{Pachucki:1996zza,Martynenko:2005rc,Carlson:2011zd} and improved in~\cite{Birse:2012eb}.  Equating the theoretical and experimental results for the Lamb shift yields~\cite{Antognini:1900ns} the previously stated result, Eq.~(\ref{eq:re}).

Alternatively, had we used the CODATA 2010 charge radius to calculate the expected Lamb shift, there would have been an $\approx 320\, \mu$eV discrepancy from the experimental value.   Either we accept the smaller radius or we must find about $320\, \mu$eV---about 10 times the value of $E_{\rm TPE}$---of further corrections.

In addition, the muonic hydrogen value of the hyperfine splitting, Eq.~(\ref{eq:hfs}), is
\begin{equation}
E_{\rm HFS}^{\rm exp} = 22.8089(51) {\rm\, meV}.
\end{equation}
This is in good accord with existing calculations of the hyperfine splitting~\cite{Faustov:2001pn,Carlson:2008ke,Carlson:2011af}.  For example, Ref.~\cite{Carlson:2011af} gives $E_{\rm HFS} = 22.8146(49) {\rm\, meV}$.  In that work, the Zemach radius is obtained from form factors fit to electron scattering experiments.  Alternatively, since the proton structure corrections are about $0.7$\% of the hyperfine splitting in muonic hydrogen (compared to about $40$ ppm in electronic hydrogen), one may fit the Zemach radius from the experimental data, to obtain
\begin{equation}
R_Z = 1.082(37) {\rm\, fm},
\end{equation}
which this time in line with the electron scattering values, albeit for this quantity there is a $3.4$\% uncertainty~\cite{Distler:2010zq}.


\subsection{Muonic deuterium}

\label{sec:deuterium}

\begin{figure}[tb]
\begin{center}

\begin{minipage}[t]{14 cm}
\centerline{  \epsfig{file=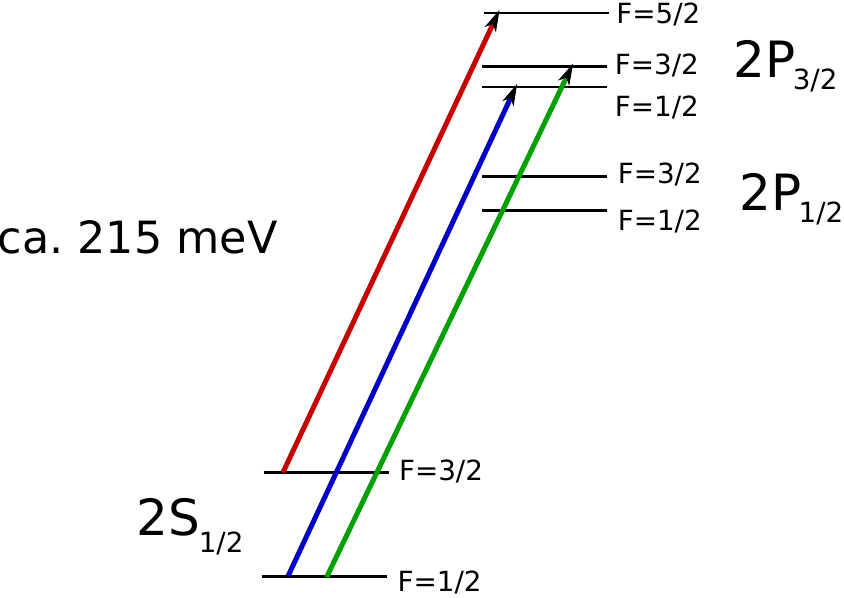,scale=1}  }
\end{minipage}

\begin{minipage}[t]{16.5 cm}
\caption{A (not to scale) diagram of the $n$\,=\,$2$ energy levels of muonic deuterium, with indication of the three level spacings that have been measured.  The $2S_{1/2,F=1/2}$ to $2P_{3/2,F=3/2}$ energy splitting is about $215$ meV.  \label{fig:deutlevel2}}
\end{minipage}

\end{center}
\end{figure}

There being a problem involving ordinary muonic hydrogen with no clear explanation, one may seek additional information about the discrepancy by studying other systems.  The CREMA group has also measured the Lamb shift in muonic deuterium.  As of this writing, the results have been discussed in open conferences but not published in a refereed journal.  Accordingly, we shall make some comments on the expectations and theory involved in this measurement,  and only a little about the actual results.  Three splittings in the muonic deuterium $2S$ state have been measured,  and are diagramed in Fig.~\ref{fig:deutlevel2}.  Since the deuteron is spin-$1$, the hyperfine splittings triple the states other than the $J=1/2$ states, and the three measured splittings are from the $2S_{1/2,F=3/2}$ to the $2P_{3/2,F=5/2}$ and from the $2S_{1/2,F=1/2}$ to both the $2P_{3/2,F=3/2}$ and $2P_{3/2,F=1/2}$ states.  The measured $2S_{1/2,F=1/2}$ to $2P_{3/2,F=3/2}$ splitting is about $215$ meV and the others are close to this value.

As a simple note, the deuteron is much larger than the proton, so the size effect is correspondingly larger.  If the radii had their CODATA 2010 values,  the proton finite size would shift the muonic $2S$ state up by $4.0$ meV, whereas the deuteron finite size would shift the same state's energy up by $27.9$ meV.  We already know the discrepancy in the proton case corresponds to lowering the $2S$ energy by about $320\,\mu$eV.  Without yet understanding the mechanism, we cannot say what the discrepancy would be in the deuteron case.  For informational purposes, one explanation (there is more discussion in a later section) involves a new particle that couples more to muons than to electrons, and also couples to protons.  If it coupled to the deuteron with the same strength as it coupled to protons (a dark photon related explanation would satisfy this criterion), then the energy discrepancy in the deuteron case would be the same except for a reduced mass correction, or about $385\,\mu$eV.   

Without committing ourselves to this or any other model, we will take $385\,\mu$eV as a benchmark expectation for the energy discrepancy.  That is, the normal deuteron size effect will be much larger than for the proton, but the discrepancy from the expected value will be about the same as for the proton.  This benchmark tells us that any uncertainties in expectations or uncertainties in theoretical corrections will have to be small compared to $385\,\mu$eV when translated into energy shifts, if they are to be useful.

A first comment concerns the direct measurement of the deuteron radius in known experiments using electrons.  One can scatter electrons, obtain a deuteron charge form factor $G_C(Q^2)$ and then obtain the deuteron charge radius.  The result, from just the scattering experiments, is
\begin{align}
R_d = 2.130 (10) {\rm\ fm},
\end{align}
or $R_d^2 = 4.537 (43)$ fm$^2$.   The uncertainty in $R_d^2$ translates into an uncertainty of about $270\,\mu$eV in the muonic deuterium $2S$ energy, so the current $e$-$d$ scattering experiments are unfortunately only of marginal use in predicting muonic deuterium energy levels to the required accuracy.  New electron-deuteron scattering experiments are underway at Mainz to significantly improve this situation~\cite{DistlerGriffioen}, and we look forward to their completion.  In the meanwhile there is another path to take.

Very accurate measurements of the $1S$-$2S$ splitting for both ordinary electronic hydrogen and deuterium are possible, and the isotope shift leads to very accurate results for the differences of the radii-squared,
\begin{align}
R_d^2 - R_p^2 = 3.82007 (65) {\rm\ fm}^2.
\end{align}
This combined with the CODATA 2010 value of the proton radius leads to the CODATA 2010 value of the deuteron radius, which is
\begin{align}
R_d = 2.1424 (21) {\rm\ fm}  ,
\end{align}
where one notes the uncertainty limit is about 5 times smaller than for the radius obtained from scattering alone.  The resulting uncertainty in the predicted deuteron size dependent energy shift in muonic deuterium is now $55\,\mu$eV, which is acceptably small.   
 
The isotope shift measurement also leads to an alternative estimate the discrepancy from the CODATA radius value that might be expected in the muonic deuterium Lamb shift.  If the isotope shift is the same for the muonic atoms as for the electronic atoms, the $320\,\mu$eV discrepancy in the proton case translates into the same discrepancy in the deuteron case, up to reduced mass effects, which is to say it leads to the same $385\,\mu$eV benchmark discrepancy mentioned above as following from a particular model.

A serious consideration in the muonic deuterium Lamb shift is the theoretical correction from the polarizability correction, or equivalently from the two-photon corrections, Fig.~\ref{fig:tpe}.  In the proton case these were important because the theory uncertainty in the calculation of these corrections~\cite{Pachucki:1996zza,Martynenko:2005rc,Carlson:2011zd,Birse:2012eb} was the largest source of uncertainty in the analysis of the measurement.  However, the actual total size of the correction was small enough that even a large error in its calculation would not affect that fact of a discrepancy between the electronic and muonic radius measurements.  The same is far from true in the deuteron case.  Inelastic contributions to the polarizability requires an energy input of only the $2.24$ MeV deuteron binding energy,  whereas a proton requires at least a pion mass of energy for a strong interaction inelastic contribution.   The size of the deuteron polarizability corrections, including the elastic parts to Fig.~\ref{fig:tpe}, is found to be of order 2 meV, and needs to be calculated to about $5$\% accuracy to allow a measurement to draw useful conclusions.

There are a number of available polarizability calculations~\cite{Friar:1978wv,Leidemann:1994vq,Friar:1997js,Pachucki:2011PhRvL.106s3007P,Friar:1997jy,Friar:2013rha} that are mainly non-relativisitic and depend on potential model considerations of the deuteron bound state.  The most recent~\cite{Pachucki:2011PhRvL.106s3007P} includes the elastic contributions and quotes an uncertainty limit well within the requirement.  In addition, much of that calculation is supported by a calculation ~\cite{Friar:2013rha} that uses the zero-range approximation which indicates that much of the result is relatively model independent.  

The result from~\cite{Pachucki:2011PhRvL.106s3007P} is
\begin{align}
\Delta E _{2P-2S}^{\rm TPE} = 1.665(16) {\rm\ meV}	\,,
\end{align}
where we have updated a small term coming from hadronic corrections from excitations of the proton and neutron, and left the uncertainty limit as it is found in~\cite{Pachucki:2011PhRvL.106s3007P}.

One would like in addition a calculation by a purely relativistic method analogous to the calculations done in~\cite{Pachucki:1996zza,Martynenko:2005rc,Carlson:2011zd} for the proton case to verify the results.  These calculations involved using data on electron proton scattering plus dispersion relations to obtain the Compton amplitudes that make up the lower part of the box in Fig.~\ref{fig:tpe}.  Changing to the deuteron case requires data on inelastic and or quasi-elastic deuteron breakup which is not well enough known in the decisive kinematic range.  The calculation has nonetheless been attempted~\cite{Carlson:2013xea}, with the result coming with a large uncertainty limit,
\begin{align}
\Delta E _{2P-2S}^{\rm TPE} = 2.01(74) {\rm\ meV}	\,.
\label{eq:cgvdh}
\end{align}
The better news here is that the same experiment that will better measure the deuteron charge radius using electron scattering will also gather useful quasi-elastic data that could help reduce the uncertainty in Eq.~(\ref{eq:cgvdh}) to a useful level~\cite{DistlerGriffioen}.


\section{Calculating the proton radius dependence}


A number of theoretical calculations and corrections are needed to convert the measured energy splittings into results for the charge radius.  The original publication, for example, gave a list of more than 20 corrections, albeit only 3 of them were large enough to affect the final result noticeable;  Ref.~\cite{Antognini:2013rsa} also contains tabulations of the calculated theoretical corrections.

There also other corrections.  A year ago one might have worried that some of the corrections were single-sourced, \textit{i.e.,} had been calculated by only one person or group.  Today, any thinkably important correction has been multiply checked; some reviews and/or additional small corrections are given in~\cite{Jentschura:2010ej,Borie:2011,Jentschura:2011ck,Pohl:2013yb}.  We shall mention here only a few topics, mainly the higher order two-photon corrections to the spectroscopy (because of its relevance to the discussion of the sources of the discrepancy), the two-photon corrections to scattering measurements (because they are still somewhat unsettled, though not apparently a decisive problem), and the lattice gauge theory proton radius results (because they may be very relevant in the future).

\subsection{Two photon corrections to the energy levels}


The two photon corrections were first calculated non-relativistically by Friar in 1979~\cite{Friar:1978wv}, whose result added to the lower order term was given as
\begin{equation}
\Delta E = \frac{2 \pi \alpha}{3} \phi(0)^2 \left( R_E^2 
	- \frac{1}{2} m_r \alpha R_{(2)}^3  \right)
\end{equation}
where $R_{(2)}^3$ is the average separation-cubed of the proton charge distribution,
\begin{equation}
R_{(2)}^3 = \int d^3 r_1 \, d^3 r_2 \   \rho_E(r_1)  \,
	|\vec r_1 -\vec r_2|^3  \rho_E(r_2)	\,
\end{equation}
and Friar has referred to $R_{(2)}$ as the third Zemach radius.

Today the calculation should be done relativistically and in momentum space, as pioneered by Pachucki~\cite{Pachucki:1999zza}.  The higher order perturbation theory calculation is represented by the diagram in Fig.~\ref{fig:tpe}, and Friar's result can be shown to be the non-relativistic limit of the elastic part of Fig.~\ref{fig:tpe}, meaning the part where the blob in the Figure is just a single proton.

An updated re-evaluation~\cite{Carlson:2011zd} has both updated this contribution and 
quantified the error limits on its numerical value,  using the latest input for what is known about the blob in the Figure.  Neither this evaluation nor other related ones obtain a result large compared to $320\,\mu$eV~\cite{Martynenko:2005rc,Birse:2012eb,Gorchtein:2013yga,Alarcon:2013cba,Peset:2014jxa}.  More detail is pushed forward to Sec.~\ref{sec:miller},  where it is needed to explain some controversy regarding a contrary claim~\cite{Miller:2011yw} that off-shell contributions existing within the blob in the Figure could indeed give contributions as large as the 320 $\mu$eV needed to resolve the proton radius problem.

\subsection{Two photon and Coulomb corrections in scattering}


There are corrections to electron-proton scattering coming from multiple photon exchange.  These are Coulomb corrections, and are dominated by two photon exchange.  The corrections at very low energies and very low momentum transfers were worked out long ago~\cite{McKinley:1948zz,Lewis:1956zz}, and at higher energies and higher momentum transfers in more modern times~\cite{Blunden:2003sp,Chen:2004tw,Afanasev:2005mp,Blunden:2005ew}.

The experimenters have made two photon corrections, basically using the older works~\cite{McKinley:1948zz,Lewis:1956zz},  which used pointlike protons.  There has been some discussion of details of the corrections~\cite{Arrington:2011kv,Bernauer:2011zza,Bernauer:2013tpr}, but the discussions did not center on the assuming of pointlike protons in the corrections. 

Recently, Lee and Milstein~\cite{Lee:2014uia} (see also~\cite{Rosenfelder:1999cd,Blunden:2005jv,Borisyuk:2006uq}) have noted that two photon corrections to the proton charge radius themselves have a correction due to the proton charge radius,  and while not claiming this extra correction is large enough to explain the proton radius puzzle, they do claim it is too large to neglect.  Including the extra correction reduces the proton charge radius measured in electron scattering by about 0.8\%,  for the lowest Mainz electron energy of 180 MeV and taking the $Q^2\to 0$ limit.  The paper~\cite{Lee:2014uia} exchanges only Coulomb photons, and it is unclear if the form factor they insert is Dirac or electric, but the basic statements appear correct.   More recently, Gorchtein~\cite{Gorchtein:2014hla} has included some relativistic corrections to the Lee-Milstein result and also given a possibly useful expression for the low-$Q^2$ contribution from inelastic intermediate states in two-photon exchange.

\subsection{Proton radius \textit{ab initio} calculations}


\textit{Ab initio} in the context of QCD includes lattice gauge theory calculations.

The current state of lattice calculations for nucleon radii and nucleon form factors is limited because lattice simulations with physical quark masses are computer time intensive and because lattice computations with disconnected quark lines are also time intensive.  One escapes the first by calculating with heavier quarks, indicating the quark mass by giving the pion mass, and escapes the second by only studying isovector nucleon quantities.
Isoscalar form factors do require disconnected diagrams, diagrams with quark loops not connected to the quark lines emanating from or ending on the lattice nucleon source or sink.  The gluons that attach the quark loops to the valence quarks are not indicated in lattice diagrams, hence the phrase ``disconnected.''   However, the disconnected diagrams contribute equally to proton and neutron, and the isovector case has been assayed.  

One may specifically focus on nucleon radii calculated from lattice gauge theory.  In the future, it may be possible and desirable to calculate using a dedicated correlator which gives directly the slope of the form factor at zero momentum transfer.  Finding such correlators by taking derivatives of known correlators is studied in~\cite{deDivitiis:2012vs} for lattice calculations of form factors that multiply Lorentz factors that go to zero at interesting points.  Applications in~\cite{deDivitiis:2012vs} are to form factors for semi-leptonic scalar meson decay, and to hadronic vacuum polarization corrections to the muon $(g-2)$.

At present, lattice calculations of nucleon radii proceed by calculating the form factor at several non-zero $Q^2$, fitting to a suitable form, typically a dipole form, and finding the radius by extrapolating to zero $Q^2$.  Results are available only for the isovector nucleon~\cite{Alexandrou:2013joa,Syritsyn2013,Green:2014xba}.  Ref.~\cite{Syritsyn2013} presents a plot of Dirac radius results for lattice calculations at various pion masses.  The results are typically about 50\% of the experimental (electron measured) isovector Dirac radii-squared, albeit rising with decreasing pion mass, with quoted uncertainties in the 10\% range.  One may say there opportunity for further work.  An uncertainty of 1\% or less for the proton alone is needed for a lattice calculation to impact the proton radius puzzle.

\section{Explanations}


If the muonic Lamb shift proton radius is correct,  three explanations have been discussed to explain the discrepancy:

1. There are unexpected QCD corrections that can give $\approx 320\,\mu$eV energy to the muonic hydrogen Lamb shift.

2. There is some ``new physics,'' a so far undiscovered new particle or particles with new interactions that violate $e$-$\mu$ universality.

3. There is a problem with the radius determined from the full collection of electronic measurements.

We will discuss each of them in turn.  Each of them may work, but each of them has problems that we shall try to mention clearly.

\subsection{``Haywire'' hadronic behavior}
\label{sec:miller}

It is possible that unexpected QCD effects could exist with unexpected magnitude.  The context of the discussion is the two-photon corrections that have already been diagramed in Fig.~\ref{fig:tpe}.  The generally accepted value has already been quoted, Eq.~(\ref{eq:tpe}), and the standard analyses are in general agreement on this value~\cite{Pachucki:1996zza,Martynenko:2005rc,Carlson:2011zd,Birse:2012eb,Gorchtein:2013yga}.  However, there are parts of the correction that are strictly speaking unknown.  The standard estimates of these parts have them extremely small, but perhaps it is otherwise~\cite{Miller:2012ne} and an examination of the issues will be given here.

We shall give some detail in order to explain how the discussion arises.  Relativistically, the corrections from two-photon exchange are obtained from the Feynman diagram Fig.~\ref{fig:tpe}.  On the scale of nuclear effects, the Fermi momenta of the atomic electrons or muons can be neglected even in a precision calculation.  The bottom part of the diagram is forward, off-shell Compton scattering.  The intermediate state in the lower leg may be a proton (the ``elastic'' case) or may be more than a proton and then ``inelastic''.  The lower part of Fig.~\ref{fig:tpe}, both elastic and inelastic, is described by the Compton tensor,  which is
\begin{align}
&T^{\mu\nu}(p,q) = \frac{i}{8\pi M}  \int d^4x	\,e^{iqx}
	\langle p | T j^\mu(x) j^\nu(0) | p \rangle  \nonumber\\[1ex]
	& \quad = \left(-g^{\mu\nu} + \frac{q^\mu q^\nu}{q^2} \right) T_1(\nu,Q^2)
	+ \frac{1}{M^2} \left( p^\mu - \frac{p\cdot q}{q^2} q^\mu \right)
	\left( p^\nu - \frac{p\cdot q}{q^2} q^\nu \right)  T_2(\nu,Q^2)		,
\label{eq:tt}
\end{align}
where $Q^2 = -q^2$ and $\nu = p\cdot q/M$ (Fig.~\ref{fig:tpe}).  Straightforwardly, the energy shift in terms of $T_1$ and $T_2$ is,
\begin{align}
\Delta E = \frac{8 M \alpha^2}{ \pi } \left|  \phi^2_n(0) \right|   \int d^4Q 
\frac{ (Q^2+2Q_0^2) T_1(iQ_0,Q^2) - (Q^2 -Q_0^2) T_2(iQ_0,Q^2) }
	{ Q^4 (Q^4 + 4M^2 Q_0^2) }		,
\end{align}
where we have done a Wick rotation so that $q_0 = iQ_0$ and $\vec Q = \vec q$.

The serious problem that we don't know what $T_1$ and $T_2$ are.  But we do have the optical theorem, that the imaginary part of a forward scattering amplitude is related to the total (elastic plus inelastic) cross section for the same incoming states.

\begin{figure}[tb]
\begin{center}
\begin{minipage}[t]{11 cm}
\centerline{ \epsfig{file=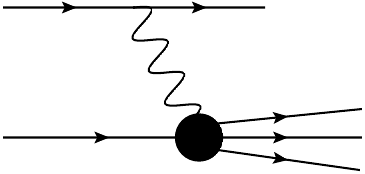,scale=1.4} }
\end{minipage}
\begin{minipage}[t]{16.5 cm}
\caption{Electron-proton scattering.\label{fig:ep}}
\end{minipage}
\end{center}
\end{figure}

In this case, the total cross section is the the one for electron-proton scattering, Fig.~\ref{fig:ep}.  The cross section over a range of kinematics has been measured at SLAC, DESY, Bonn, JLab, Mainz, and other laboratories and codified in the spin averaged case in terms of two structure functions $F_1(\nu,Q^2)$ and $F_2(\nu,Q^2)$.  These then give the imaginary parts of the Compton amplitudes as
\begin{align}
{\rm Im\,} T_1(\nu,Q^2) &= \frac{1}{4M} F_1(\nu,Q^2) 	,	\nonumber\\
	{\rm Im\,} T_2(\nu,Q^2) &= \frac{1}{4\nu} F_2(\nu,Q^2)
\end{align}
(the normalization of the $F_i$ is standard, the normalization of the $T_i$ follows Eq.~(\ref{eq:tt})).  From the imaginary part of the $T_i$ and the Cauchy integral formula, leading to what are known as dispersion relations, we can obtain the full amplitudes.

But there is a glitch, and this glitch underlies the present discussion.  When using the Cauchy integral formula, we are integrating in $\nu$ and some of the integrals do not converge at high $\nu$.  This affects $T_1$ but not $T_2$.  So for $T_2$ we can write a simple dispersion relation, but for $T_1$ we instead write a dispersion relation for $T_1/\nu$.  The convergence is now fine, but now we have a pole at $\nu = 0$,  and the price is that we need to know $T_1(0,Q^2)$.  The dispersion relations are
\begin{equation}
T_1(q_0,Q^2) =  T_1(0,Q^2)  + \frac{q_0^2}{2\pi M} \int_{\nu_{0}}^\infty  d\nu 
	\frac{F_1(\nu,Q^2)}{\nu (\nu^2 - q_0^2) }		\,,
\end{equation}
and
\begin{align}
T_2(q_0,Q^2) =   \frac{1}{2\pi} \int_{\nu_{0}}^\infty  d\nu 
	\frac{ F_2(\nu,Q^2) }{ \nu^2 - q_0^2 }		\,.
\end{align}
The dispersion relation for $T_1$ is a subtracted dispersion relation and the term $T_1(0,Q^2)$ is the ``subtraction function.''   

To get the energy shift, there is still an integral needed over $Q^2$.  Hence we need to know $T_1(0,Q^2)$ for all non-negative $Q^2$.  This requires modeling because only the point $Q^2 = 0$ corresponds to physical kinematics.  The slope in $Q^2$ of $T(0,Q^2)$ at $Q^2=0$ can be obtained from low energy expansions~\cite{Pachucki:1996zza,Carlson:2011zd},  and more recently the curvature at $Q^2=0$ has been obtained using chiral perturbation theory~\cite{Birse:2012eb}.  Additionally, the scaling behavior at high-$Q^2$ is fixed from perturbative QCD considerations~\cite{Hill:2011wy}.  Hence it is possible to make a reasonable, to most workers, treatment of the subtraction function and assign it a reliable uncertainty limit.  Inserting inelastic and elastic experimental knowledge of $F_1$ and $F_2$ for other terms in the integrals is straightforward.  Using up to date data~\cite{Carlson:2011zd} and treatment of the subtraction function from~\cite{Birse:2012eb} one has $\left(-4.2(1.2) + 12.7(0.5) + 29.5(1.3)\right)\, \mu$eV from the subtraction term, the inelastic contributions from $F_{1,2}$, and the elastic contributions from $F_{1,2}$, respectively, summing to the result quoted earlier in Eq.~(\ref{eq:tpe}).  The result is noticeable (and, incidentally, gives a large fraction of the total uncertainty quoted for the muonic Lamb shift measurement), but it is not the $320\,\mu$eV needed to explain the discrepancy while keeping the CODATA proton radius

However, it is still true that the subtraction function is not really known.  It is possible to speculate on forms for $T_1(0,Q^2)$ that are innocuous for low $Q^2$ but have a large peak at (say) $Q^2 \gtrsim 10$ GeV$^2$ before settling down to the correct asymptotic behavior.  It is possible to thereby obtain $320\, \mu$eV for the subtraction term, and so ``solve'' the proton radius problem as has been done in~\cite{Miller:2012ne}.

\begin{figure}[tb]
\begin{center}
\begin{minipage}[t]{11 cm}
\centerline{ \epsfig{file=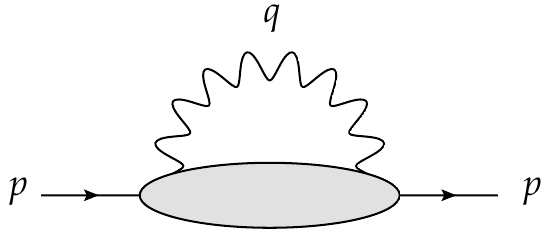,scale=1.4} }
\end{minipage}
\begin{minipage}[t]{16.5 cm}
\caption{Electromagnetic contribution to nucleon mass.\label{fig:cottingham}}
\end{minipage}
\end{center}
\end{figure}

But the modification of the subtraction function does have further consequences.  One should think about the Cottingham formula~\cite{Cottingham:1963zz,Gasser:1974wd,WalkerLoud:2012bg}, which gives the electromagnetic contribution to the neutron-proton mass difference. The Compton tensor also enters here, illustrated in Fig.~\ref{fig:cottingham}, where the two photon lines that attach to the nucleon are tied together rather than attached to a lepton.  One has the same subtraction function.  The published subtraction function proposed to solve the proton radius problem contributes a startling $600$ MeV to the individual nucleon terms in the Cottingham formula.  The measured neutron-proton mass difference is $1.29$ Mev~\cite{Mohr:2012tt}.  Even if one makes the large term in the subtraction function isoscalar, so that is does not affect the mass difference, one is still at this moment left with an implausible $600$ MeV electromagnetic contribution to the individual nucleon masses.    

The validity of perturbative QED for electron-proton interactions has also been questioned~\cite{Jentschura:2014ila,Pachucki:2014zea}, but these explanations would seem to require more explanation of how the validity of perturbative QED can be vitiated in interactions with hadronic bound states.


\subsection{Exotic Explanations}


A possibility is that the discrepancy is a signal of new physics, a new phenomenon, perhaps an exchange of a new particle, that we are uncovering for the first time.  See~\cite{TuckerSmith:2010ra,Barger:2010aj,Barger:2011mt,Batell:2011qq,Brax:2010gp,Jentschura:2010ha,Carlson:2012pc,Wang:2013fma,Onofrio:2013fea,Karshenboim:2014tka}.  Some of the references discuss new physics possibilities, some discuss problems that ensue, some discuss both, and some suggest possible further experiments. 

Ref.~\cite{Jaeckel:2010xx} shows that a new attractive (the $320\,\mu$eV shift being downwards) particle that interacts equally with electrons and muons would in fact lead to a larger radius from muonic hydrogen, the opposite of what is observed.   Hence we need to consider a breakdown of muon-electron universality.  A difference between the interactions of the muon and electron, other than the obvious gravitational one, has been long sought and not found (some early hints~\cite{Camilleri:1969ag,Braunstein:1972st} were considered ``not experimentally significant''~\cite{Braunstein:1972st} even in the original sources).   The overall negative result of these searches will lead to difficulty in the present context.

In particular, enhanced coupling to muons necessarily affects the theory for $(g-2)_\mu$, where there is in fact a known discrepancy, albeit one that is small in percentage terms compared to the proton radius discrepancy.  However, without fine tunings, the effect of any new particle that can explain the proton radius puzzle will have a large effect upon $(g-2)_\mu$.  Hence there is a sticking coupling:  one cannot claim the theory corrections to $(g-2)_\mu$ are under good control unless the proton radius puzzle is understood.

Considering the workings of a new particle in a way that is common to (at least) Refs.~\cite{TuckerSmith:2010ra,Batell:2011qq,Carlson:2012pc}, consider a particle that couples to muon and protons, but with no coupling or weaker coupling to electrons.  The particlecan be scalar or vector.  Pseudoscalars or axial vectors give small contributions at low momentum transfers, \textit{i.e.}, to atomic binding, for similar couplings.

\begin{figure}[tb]
\begin{center}
\begin{minipage}[t]{11 cm}
\centerline{ \epsfig{file=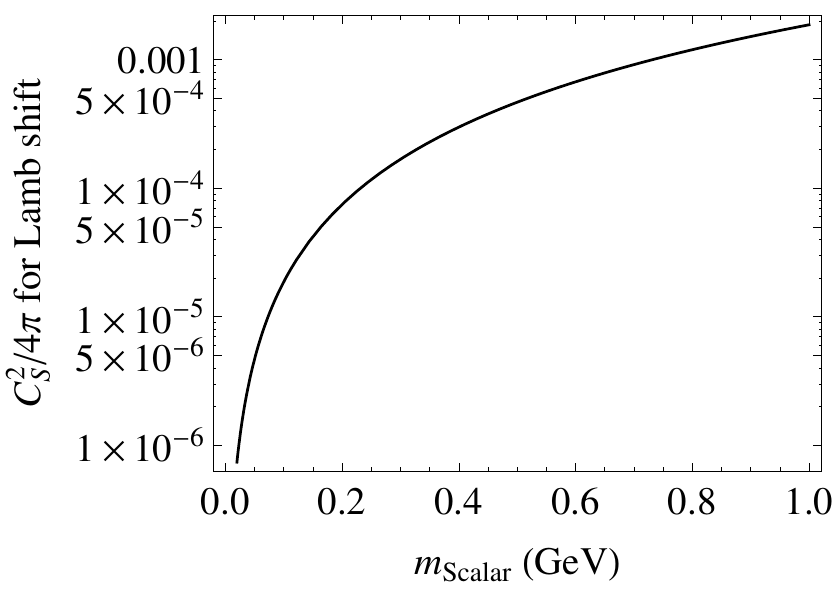,scale=1.0} }
\end{minipage}
\begin{minipage}[t]{16.5 cm}
\caption{New particle scalar couplings vs. mass.\label{fig:cvsm}}
\end{minipage}
\end{center}
\end{figure}

For illustration, the scalar case has an interaction Lagrangian
\begin{equation}
{\mathcal L}_S =  - C^\mu_S \ \bar \mu \ \phi \ \mu - C^p_S \ \bar p \ \phi \ p	\,,
\end{equation}
where muon and proton fields are denoted by the symbol for the particle and $\phi$ is a scalar field, and $C^{\mu,p}_S$ are constant coupling parameters.  This gives a non-relativistically calculated $2S$ -$2P$ energy shift
\begin{equation}
\Delta E = \int r^2 dr \ V(r) \left( R_{20}^2 - R_{21}^2 \right)
	=  - \frac{C_S^\mu C_S^p}{4\pi} 
	\frac{M^2 (m_r\alpha)^3}{2(M+m_r\alpha)^4}		\,,
\end{equation}
where $M$ is the mass of the scalar $\phi$ and $m_r = m_\mu m_p/(m_\mu+m_p)$.  The $\phi$ mass fixes the product of the couplings, since we must obtain $\Delta E = 320\,\mu$eV.  For the case $C_S^\mu = C_S^p =C_S$, the result is shown graphically in Fig.~\ref{fig:cvsm}.  A plot for the vector case would look the same.

Now for $a_\mu = (g-2)_\mu/2$,   the discrepancy between theory and experiment is 2 part-per-million (ppm) using only standard physics.  In more detail, the results for $a_\mu$ are
\begin{align}
a_\mu({\rm data}) &= (116\, 592\, 089 \pm 63) \times 10^{-11} \quad [0.5 \rm{\ ppm}],
		\nonumber\\
a_\mu({\rm thy.}) &= (116\, 591\, 840 \pm 59) \times 10^{-11} \quad [0.5 \rm{\ ppm}],
		\nonumber\\
\delta a_\mu &= (249 \pm 87) \times 10^{-11} \quad 
		[2.1 \rm{\ ppm} \pm 0.7 \rm{\ ppm}].
\end{align}
(As a note, the above is from~\cite{Aoyama:2012wk}, where the completed 5-loop QED $(g-2)$ calculations were reported.  Their theory number used the hadronic corrections from Hagiwara {\textit et al.}~\cite{Hagiwara:2011af}, which was one of three papers reporting hadronic corrections to $(g-2)_\mu$ in 2011.  Hagiwara {\textit et al.} was the largest, and in units of $10^{-11}$, Davier {\textit et al.}~\cite{Davier:2010nc} were 20 units smaller, and Jegerlehner and Szafron~\cite{Jegerlehner:2011ti} were about 20 units smaller still.  Hence the discrepancy could be about 40 units larger than shown above.)

\begin{figure}[tb]
\begin{center}
\begin{minipage}[t]{8 cm}
\epsfig{file=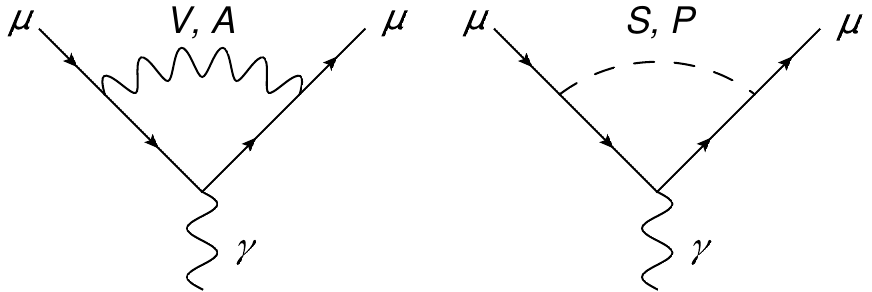,scale=1.0}
\end{minipage}
\begin{minipage}[t]{16.5 cm}
\caption{Corrections to the muon magnetic moment due to new particle exchange\label{fig:gminustwo}}
  \end{minipage}
\end{center}
\end{figure}

What makes it possible to have small corrections to $(g-2)_\mu$ is the fact that corrections to the anomalous magnetic moment, Fig.~\ref{fig:gminustwo}, from scalar and pseudoscalar particles have opposite signs.  The same is true for corrections from polar vector and axial vector exchanges.

\begin{figure}[tb]
\begin{center}
\begin{minipage}[t]{11 cm}
\centerline{ \epsfig{file=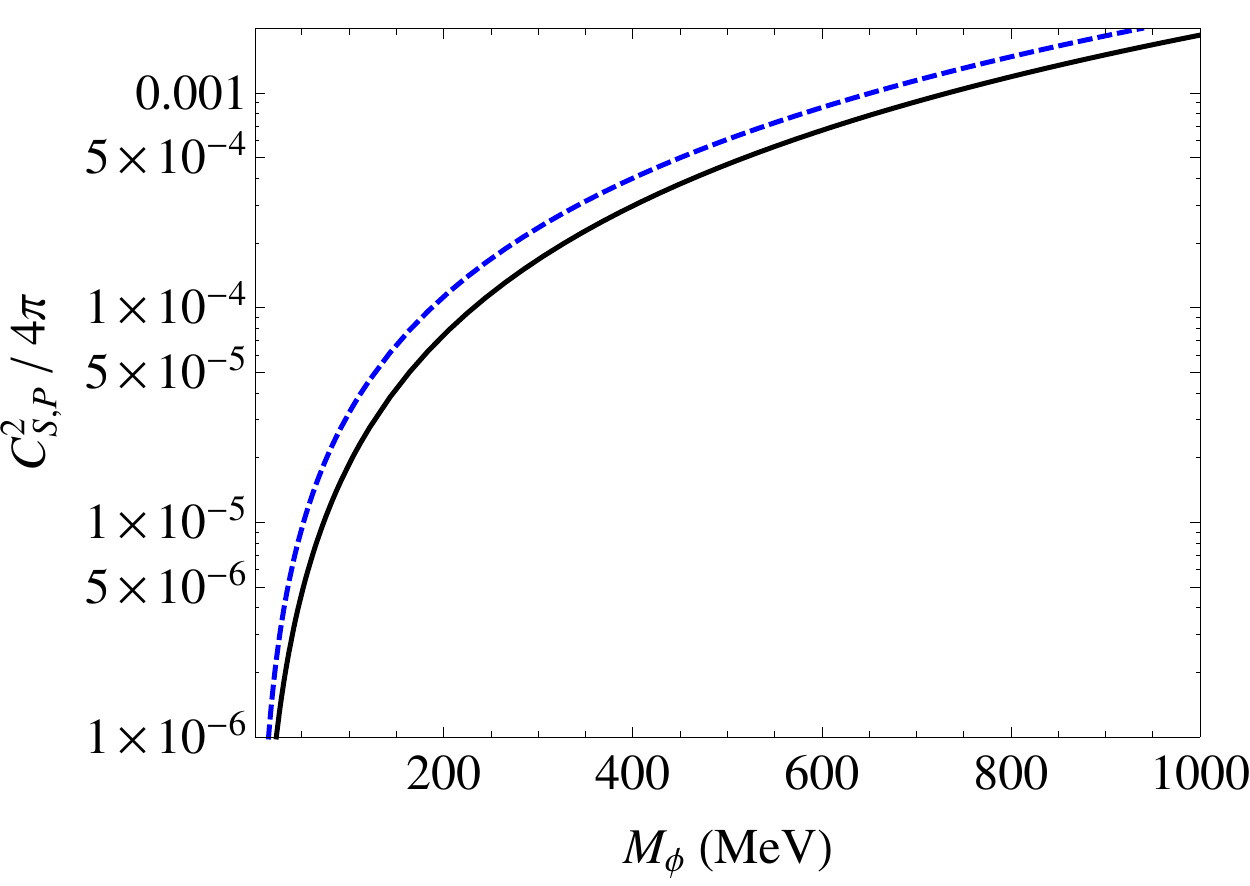,scale=0.70}
 \epsfig{file=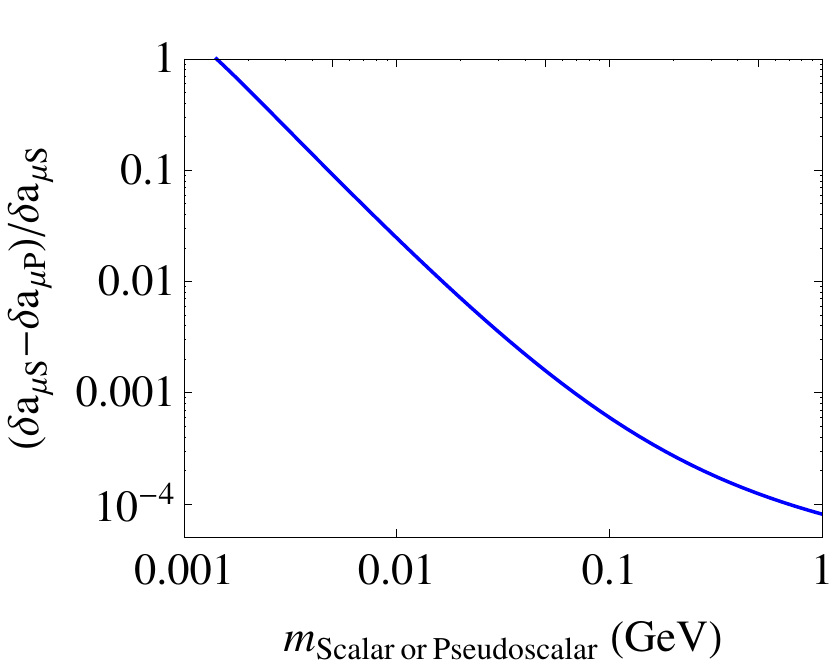,scale=0.97 } }
 \vskip 4 mm
 \centerline{  \epsfig{file=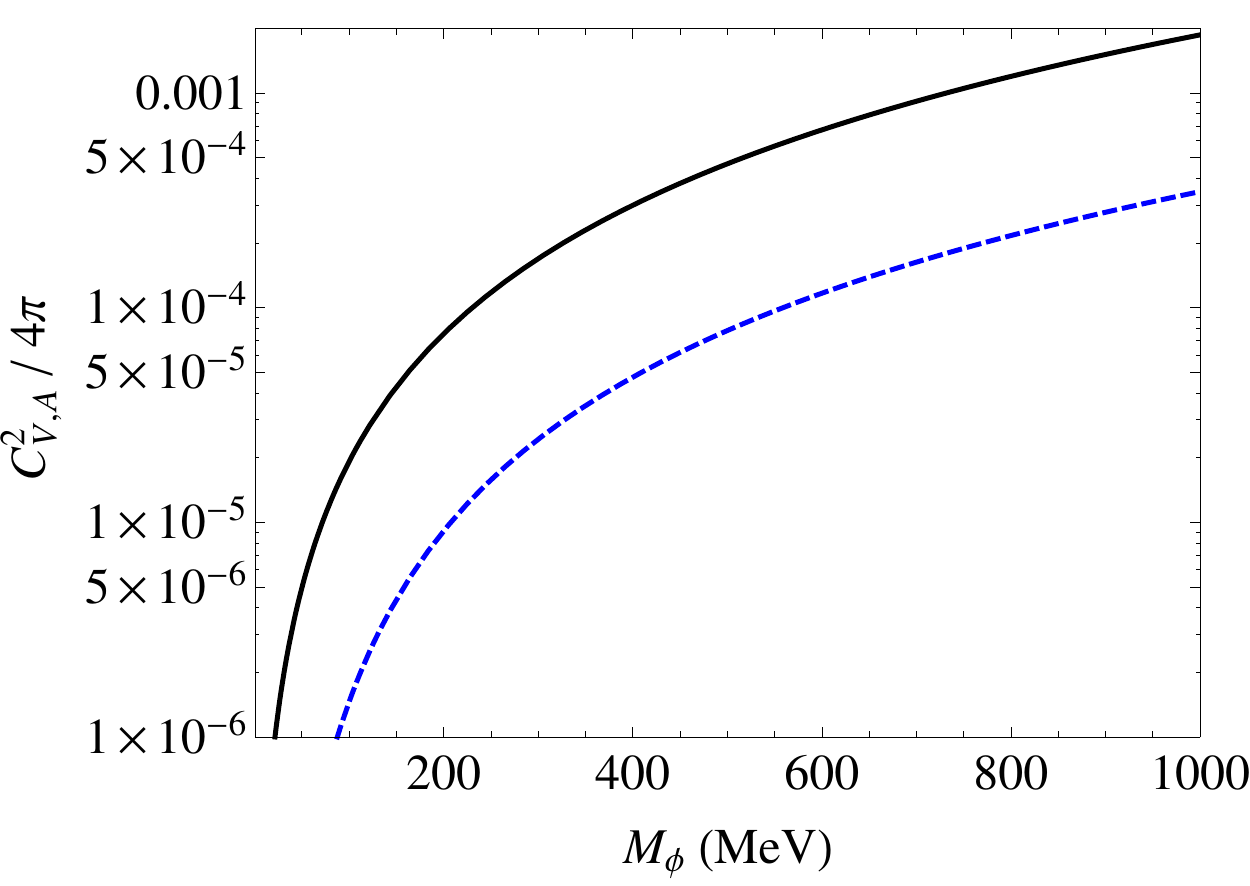,scale=0.70 }  }
\end{minipage}
\begin{minipage}[t]{16.5 cm}
\caption{Top left panel: New physics $S$, $P$ couplings required to both explain the proton radius problem and fine tune $(g-2)_\mu$.  The solid line is $C_S$ and the dashed line is $C_P$.  Top right panel: Degree of fine tuning required for the scalar-pseudoscalar case. Bottom panel:  New physics $V$, $A$ couplings required to both explain the proton radius problem and fine tune $(g-2)_\mu$.  The solid line is $C_V$ and $C_A$ is dashed. \label{fig:fine-tune}}
\end{minipage}
\end{center}
\end{figure}

If, as an example, we include both new scalar and new pseudoscalar particles, we have the amended Lagrangian,
\begin{equation}
{\mathcal L}_S =  - C^\mu_S \, \phi_S \, \bar \mu\, \mu - C^p_S \, \phi_S \, \bar p  \, p
- iC^\mu_P\, \phi_P \,  \bar{\psi}_\mu \gamma_5 \psi_\mu
- iC^p_P \, \phi \, \bar{\psi}_p  \gamma_5 \psi_p	\,.
\end{equation}
For a given mass and the already determined $C_S$, one can find the $C_P$ required to keep $\delta a_\mu$ at or below the known discrepancy.  The result is shown in the left hand panel of Fig.~\ref{fig:fine-tune}.  Another way of expressing the result is in the degree of fine tuning required.  The difference between the individual contributions to the anomalous magnetic moment from the scalar and pseudoscalar exchanges, compared to the contribution from either one of them, is shown in the center panel of Fig.~\ref{fig:fine-tune}.  The fine tuning becomes extreme as the mass of the exchanged particle rises.  However, at low masses the contribution of the scalar by itself to $(g-2)_\mu$ is small enough that no cancellation from a pseudoscalar is needed.  This in fact is the situation envisioned in~\cite{TuckerSmith:2010ra}.  A plot showing the couplings needed for the vector-axial vector case is shown in the right-hand panel.  Fine tuning is also required here, but not to the degree seen in the scalar-pseudoscalar case.

\subsubsection{Constraints from $K$-decay}

\begin{figure}[tb]
\begin{center}
\begin{minipage}[t]{11 cm}
\centerline{ \epsfig{file=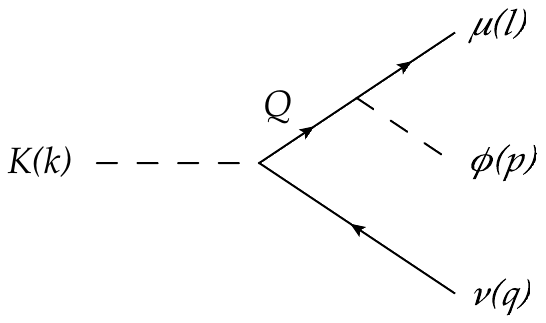,scale=1.0} }
\end{minipage}
\begin{minipage}[t]{16.5 cm}
\caption{Kaon decay with an extra neutral scalar, $\phi$, either scalar or pseudoscalar. \label{fig:kdk}}
\end{minipage}
\end{center}
\end{figure}

New particles that couple to muons can give radiative corrections to any decay involving muons, and an immediate example is radiative corrections to $K^\pm \to \mu^\pm \nu$, Fig.~\ref{fig:kdk}.  Depending on the coupling to electrons, the $\phi$ particle in Fig.~\ref{fig:kdk} can either exit the detector~\cite{Pang:1989ut,Barger:2011mt,Carlson:2012pc}, or decay into $e^+ e^-$ pairs~\cite{Carlson:2013mya}.  Both possibilities are interesting.

If the $\phi$ leaves the detector without decay or interaction, the only visible decay particle will be a muon, with energy below the energy it would have in the two body decay  $K^\pm \to \mu^\pm \nu$.  Experiments have hunted for $K^\pm \to \mu^\pm + {\rm\, invisible}$~\cite{Pang:1989ut},  where ``invisible'' means particles that are neutral but not photons.  The original motivation appears to have to explore the possibility that there were multiple neutrinos in the final state, but the bound applies to the present case also.  The results lead to mass constraints that eliminate low mass ($m_\phi < 160 MeV$) new vector particles or eliminate middle mass ($90 < m_\phi < 200$ MeV) new scalars~\cite{Barger:2011mt,Carlson:2012pc}.

If the $\phi$ can decay into $e^+ e^-$,  the bounds of the last paragraph don't apply.  The decay that does occur, $K\to \mu \nu \phi \to \mu \nu e^+ e^-$, competes with the standard QED radiative decay $K\to \mu \nu \gamma^* \to \mu \nu e^+ e^-$~\cite{Bijnens:1992en,Krishna:1972za}.   The former case of course has the $e^+ e^-$ mass concentrated at the mass of the $\phi$ and with a mass choice, the couplings to the muon are fixed.  The $\phi$ is certain to decay in the detector unless the electron couplings are very small, and theory studies show a definite $e^+e^-$ peak above background~\cite{Carlson:2013mya}, if the new particle in fact exists.

\subsubsection{HFS and other constraints}

Ref.~\cite{Karshenboim:2014tka} points out that the new particle exchange can contribute to the HFS in muoninum [the $\mu e$ bound state], which is accurately measured and for which the standard model theory is fairly accurate .  Hence a small new contribution can matter, and and new particle that couples to muons can still couple to electrons through a muon loop that attaches to photons that carry the interaction to the electrons.  The size of the contribution requires the new particle mass to lie under $25$ MeV.

Further, the measurement of the HFS in muonic hydrogen~\cite{Antognini:1900ns},
\be
\Delta E_{\rm HFS}^{\rm exp} = 22.8089(51) {\rm\ meV},
\ee
agrees within uncertainties with the value calculated using known physics~\cite{Carlson:2008ke}.  This also places pressure on exotic proton radius puzzle explanations.  Pseudoscalar or axial particles were introduced for the fine tuning of $(g-2)_\mu$ and these particles do not contribute strongly to the overall muonic hydrogen energy shift.  But they do have in the nonrelativistic limit a noticeable spin-spin force and hence affect the HFS predictions.  We will give a few details.  To state the conclusion at the outset, lower mass exotic exchanges are still possible, high mass exotic exchanged particles are not.

\begin{figure}[h]
\begin{center}
\begin{minipage}[t]{11 cm}
\centerline{ \epsfig{file=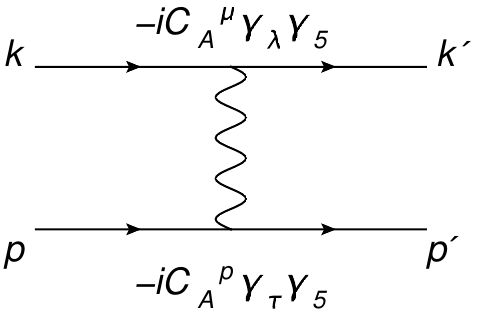,scale=1.0} }
\end{minipage}
\begin{minipage}[t]{16.5 cm}
\caption{Axial vector exchange in an atomic system.\label{fig:axialexch}}
\end{minipage}
\end{center}
\end{figure}

\begin{figure}[tb]
\begin{center}
\begin{minipage}[bt]{17 cm}
{ \epsfig{file=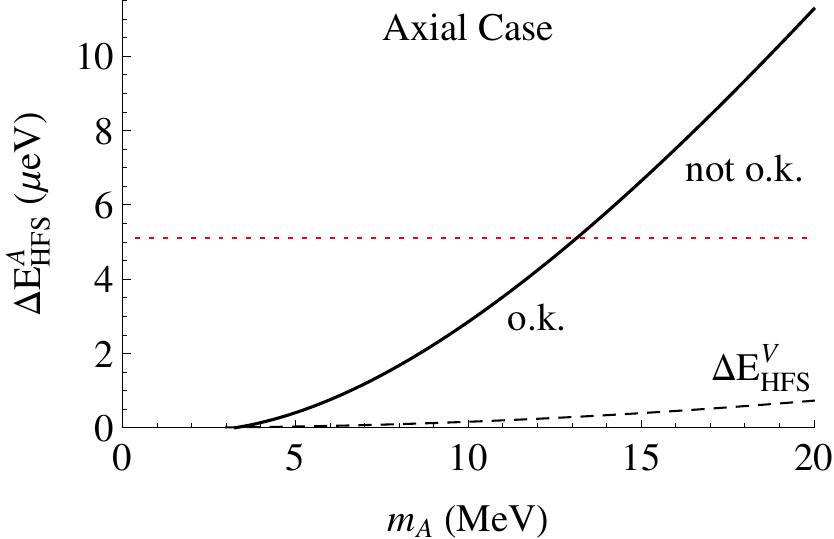,scale=0.9}  \hfil
		\epsfig{file=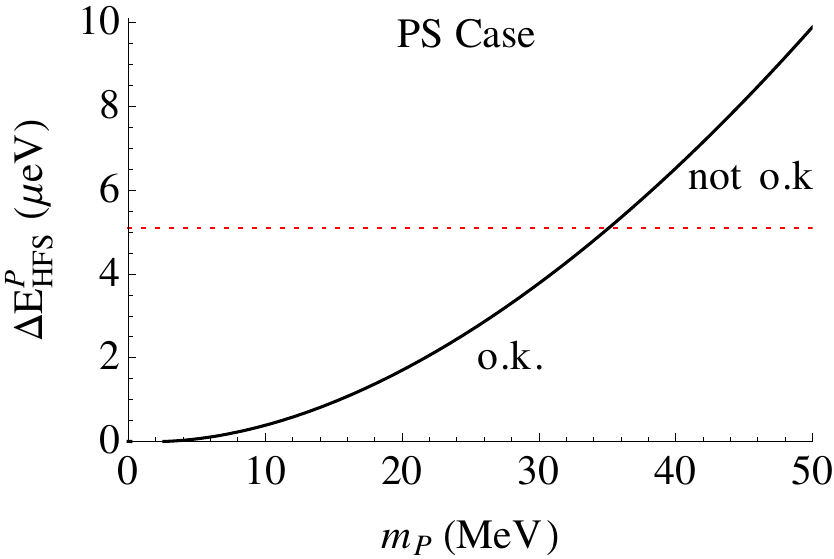,scale=0.9}  }
\end{minipage}
\begin{minipage}[t]{16.5 cm}
\caption{Contributions of exotic particle exchanges (in one model~\cite{Carlson:2012pc}) where the exotic particle is axial vector (left panel) or pseudoscalar (right panel).  To avid conflict with data and with standard model calculations, the contribution should be below $5.1\ \mu$eV, indicated by a the dotted line. \label{fig:hfs_constraint}}
\end{minipage}
\end{center}
\end{figure}

For definiteness, consider the axial exchange in Fig.~\ref{fig:axialexch}.  The calculation is similar to one of Drell and Sullivan~\cite{Drell:1965is} in another context.  In the nonrelativistic limit the matrix element is,
\begin{align}
\mathcal M_A = 2m_\mu \, 2m_p \frac{C_A^\mu C_A^p}{\vec q^2 +m_A^2} \ 
	\vec \sigma_\mu{\cdot} \vec \sigma_p		\,,
\end{align}
and the contribution to the HFS for the $2S$ state is
\begin{align}
\Delta E^A_{\rm HFS} = - \frac{C_A^\mu C_A^p}{4\pi} \ \frac{2 m_r^3 \alpha^3}{m_A^2} \ 
	\frac{m_A^2 (m_A^2 + \frac{1}{2} m_r^2 \alpha^2) }{ (m_A+m_r \alpha )^4}	.
\end{align}
With
${C_A^\mu C_A^p}$ as a function of the exchanged mass $m_A$ already obtained from the fine tuning constraints, there follows the HFS results \textit{vs.}~mass shown in Fig.~\ref{fig:hfs_constraint}, left panel.  To obtain a contribution to the HFS that is below the allowed tolerance, the axial vector mass should be below about $13$ MeV.  (The corresponding result for the polar vector exchange is also shown on the plot.)  This, of course, has been done in one particular model~\cite{CR:2014}, but other exotic models explaining the proton radius puzzle will behave qualitatively similarly.

A similar plot for the pseudoscalar case appears in Fig.~\ref{fig:hfs_constraint}, right panel.  In this case the allowed mass can be somewhat higher, but still needs to be below about $35$ MeV.

Another series of constraints comes from searches for decays $\Upsilon \to \gamma + {\rm\, invisible}$ or $J/\psi \to \gamma + {\rm\, invisible}$, which will emphasize the targeted couplings of any putative new particles.  The first of these is a $b\bar b$ bound state, and the other a $c\bar c$ bound state. The motivation for these searches was to find glueballs.  A large part of the decay at the perturbative level should be into $\gamma g g$ final states, where $g$ is a gluon, and it was hoped that the $g g$ would often bind into glueball state with a definite mass an hence bumps in the $\gamma$ spectrum.  No such states were found.  The data left behind is a serious constraint here, because the couplings for a given mass are now known for the proton and if the coupling were the same for the $b$ or $c$ quark, the new particles that fix the proton radius problem would be seen here~\cite{Barger:2010aj}.  This then leads to forbidding coupling of the new particle to second or third generation quarks.

Finally, Ref.~\cite{Karshenboim:2014tka} emphasizes the need for a gauge-invariant, renormalizable, theory for the new physics explanations.  The interactions that have been written should be viewed as low energy approximations to a complete theory.  A non-renormalizable theory has interactions whose effect grows excessively with energy, and in the present case the high energy problems are visible in the large radiative corrections to $W\to \mu\nu$ decay~\cite{Karshenboim:2014tka}.

Alltogether, the prospect of new particles explaining the proton radius problem is difficult but not impossible.  Low masses are favored.  One can to``explain" two vexing problems of modern physics, the proton radius problem and the muon anomalous magnetic moment problem, using the existence of several possible particles with unusual properties.  The price is high.  There need to be targeted couplings and fine tuning.  Nonetheless, the idea is not eliminated.


\subsection{Electronic scattering measurements}


There are many experiments using electrons that determine the proton charge radius, and only one using muons.  So it is natural to focus on the muon experiment, but one could also entertain the idea that the charge radius determined using electrons is not as secure as the error limits suggest.  We begin by discussing the radius obtained from electron scattering and list some alternating analyses.

One has begin by respecting the fit of the experimenters themselves to their own data~\cite{Simon:1980hu,Bernauer:2010wm,Bernauer:2013tpr,Zhan:2011ji}.  The quality and number of electron scattering data from the 2010 Mainz experiment~\cite{Bernauer:2010wm} is remarkable.   They have 1422 data points, with $Q^2$ from about 0.004 GeV$^2$ to about 1 GeV$^2$.  They fit their entire data set with a number of different parameterizations of the electric and magnetic form factors,  and this is where one obtains the electric radius  quoted earlier.

However, the dispersive analysis reported in~\cite{Lorenz:2012tm} also uses only electron scattering data but obtains a charge radius in accord with the $\mu$-H value.  They treat the form factors as analytic functions, and there are cuts that show that fits to the data using functions like polynomials in $Q^2$ are not guaranteed convergent for $Q^2$ above $4m_\pi^2$ (on this point see also~\cite{Hill:2010yb,Lorenz:2014vha}).  They extend the analysis to include timelike data, which can done using the analyticity properties of the scattering amplitudes and the dispersion theory.    One viewpoint on the dispersive treatment is that it gives fit functions that satisfy the necessary analyticity requirements at all $Q^2$ an hence are convergent.  The results of~\cite{Lorenz:2012tm} give a proton charge radius $R_E = 0.84(1)$ fm.

On the other hand, the authors of~\cite{Hill:2010yb} also respect the analytic properties of the form factors, but they find a result well in accord with the CODATA value.  They conformally map $Q^2$ onto another variable, where the Taylor expansion in the other variable is expected to be convergent everywhere.  They fit the data catalogued in~\cite{Rosenfelder:1999cd}, which clearly does not include the newest data, and find stable fits with $R_E = 0.870(26)$ fm (combining in quadrature their two quoted uncertainties).  An update of this work including all currently available data is underway~\cite{Hill:2014}.  However, currently available is similar work using the conformally mapped variable done by~\cite{Lorenz:2014vha}, fitting the entire Mainz data set and finding the lower $R_E = 0.84(1)$ fm proton radius.  A difference between~\cite{Hill:2010yb} and~\cite{Lorenz:2014vha}, in addition to the different data sets, is that the former restricted the size of the coefficients in the expansion, whereas the latter did not.  The former may argue that the coefficients are thereby a physically more reasonable size, the latter may point out the significantly lower $\chi^2$ per degree of freedom obtained.

Going in another direction, the charge radius is a low $Q^2$ quantity, so that low $Q^2$ data can suffice to determine it.  The determination requires finding the difference of the form factor from unity at small $Q^2$, but the measurement range should not be too small lest uncertainties in the measurement make determination of the slope impossible.  On the other hand, for a data selection covering a wide range of $Q^2$, the form factors measured at the high part of the range   have little relevance to determining the slope at $Q^2 = 0$.  Ref.~\cite{Sick:2014PhRvC..89a2201S} suggests that the range of $Q^2$ needed to determine the slope is in the range $0.01$ to $0.06$ GeV$^2$.

Considering this first over a somewhat narrower range, of the 1422 Mainz 2010 data points, about 200 are at $Q^2$ below $0.02$ GeV$^2$, and have small uncertainty.   Fitting just these circa 200 data points leads strikingly to a radius agreeing with the $\mu$-H value with about $2\%$ uncertainty [$R_E = 0.842 \,(20)$ fm].  An additional detail is that the fit has terms up to $(Q^2)^2$, terms beyond that being not determinable from and by expectation not important in low $Q^2$ data.  The curvature, the coefficient of the $Q^4$ term, obtained has small central value but large uncertainty limit.  The attitude one takes to this depends on whether one focuses on the former or the latter~\cite{Griffioen2014}.  The result that the radius comes out to the lower value continues even as the upper limit of $Q^2$ in the data selection doubles or triples~\cite{Griffioen2014}.

Also finding large radii are~\cite{Sick:2014PhRvC..89a2201S,Sick:2012zz,Sick:2003gm}.  Ref.~\cite{Sick:2014PhRvC..89a2201S} points out that the proton matter density at large $r$ might give an effect on the form factor that is small overall, but large at very low $Q^2$ and vitiate the extrapolation to $Q^2=0$.  (An similar point was pursued in~\cite{derujula:2011PhLB..697...26D}, but with an implementation giving form factors far from data at measured $Q^2$~\cite{Cloet:2011PhRvC..83a2201C,Distler:2010zq}.)  They also, perhaps not in concert with the first point, advocate including high $Q^2$ data when fitting the charge radius.  In any case, fits are done with continued fractions~\cite{Sick:2003gm} and sum of gaussians forms~\cite{Sick:2012zz} (including the Mainz data), with a result in the more recent work $R_E = 0.886(8)$ fm.

Perhaps the only sure conclusion from these conflicting claims is that the uncertainty in the proton radius obtained from electron scattering is larger than what is quoted.  The large number of recent analyses on related issues illustrates that the precise fitting of the proton form factor in
the low $Q^2$ region is not straightforward and remains a matter of ongoing discussion.  Data at still lower $Q^2$ and muon scattering data would be helpful, and there are such plans for both $e$-$p$ and $\mu$-$p$ elastic scattering, which will be mentioned in separate sections below.  

Of course, electronic measurements of the proton radius come from atomic spectroscopy as well as from electron scattering.  Again, no single measurement gives the quoted better than 1\% accuracy, but rather there are on the order of 15 measurements with accuracies of a few percent that legitimately give the smaller error limit when combined.  If one wants to entertain the idea that the proton radius is really the lower muonic Lamb shift value, one will have to explain why the spectroscopy results give the larger value.  However, there are no known problems with the atomic spectroscopy measurements and no alternative analyses obtaining different radii.  


\section{The Future}


The proton radius puzzle remains unsolved.  Further, it has implications for other fields.  Since exotic explanations must impact other purely muonic processes, the theory for the $(g-2)_\mu$ measurements cannot be deemed under control until the proton radius puzzle is solved~\cite{Karshenboim:2014tka}.

There are a number of experiments relevant to the proton radius puzzle, and a small and perhaps incomplete catalog is given here, along with some rather brief comments about what the experiment is and why it connects to this puzzle.  It is exciting to anticipate the large amount of potentially useful information that is coming.


\subsection{New CREMA measurements}


The published CREMA measurements are for muonic hydrogen~\cite{Pohl:2010zz,Antognini:1900ns}.  The collaboration also measured  2S-2P transitions in muonic deuterium simultaneously with the first muonic hydrogen run, which are presently being analyzed, and a similar measurement of the Lamb shift in muonic helium ions has been performed more recently at the PSI.   A combined analysis of these experiments will help to clarify whether the problem arises in the muonic sector, or if it is related to the proton, or the Rydberg constant, or if it originates from bound-state QED,  or if it is something still more startling.


\subsection{MUSE}
\label{ssec:muse}


The MUon proton Scattering Experiment (MUSE)~\cite{Gilman:2013eiv} at the Paul Scherrer Institute is a simultaneous measurement of $\mu^+ p$ and $e^+ p$ elastic scattering, as well as of $\mu^- p$ and $e^- p$ elastic scattering.  The main interest here is in filling in a category of measurement.  That is, the proton radius has been measured consistently using electrons in both scattering and spectroscopy;  it has been measured using muons only via spectroscopy.  This experiment will be the first see if muon-proton scattering gives the same answer as muonic hydrogen spectroscopy.
 
 In addition, the differences between positive and negative charge lepton scattering measure two-photon exchange effects, which are higher-order corrections to the scattering process. The experimenters plan to determine the proton radius from $\mu p$ scattering at the level of about  $0.01$ fm, similar to previous $e$-$p$ measurements.  They hope to measure momentum transfers in the range $0.002$ to $0.07$ GeV$^2$.  There have been some test runs already, with two 6-month production runs scheduled for 2016 and 2017.


\subsection{PRad}


This experiment will measure $e$-$p$ elastic cross sections at high precision and at very low four-momentum transfer squared, $Q^2$, from about $2\times 10^{-4}$ to $2\times 10^{-2}$ GeV$^2$, in Hall B at Jefferson Lab~\cite{Gasparian:2014rna}.  They obtain the low $Q^2$ by having low scattering angles, which is possible because they have no bulky magnetic spectrometers, but just a target and a downstream lead-glass wall acting as a calorimeter.    The absolute value of the $e$-$p$ cross sections is calibrated by a well known QED process, M{\o}ller scattering or $e^-$-$e^-$ elastic scattering, which will be continuously measured in this experiment with similar kinematics and the same apparatus. With accurate data and low $Q^2$, they hope to obtain a sub-percent and essentially model independent extraction of the proton charge radius.


\subsection{ISR form factor measurements}

\begin{figure}[tb]
\begin{center}
\begin{minipage}[t]{8 cm}
\centerline{  \epsfig{file=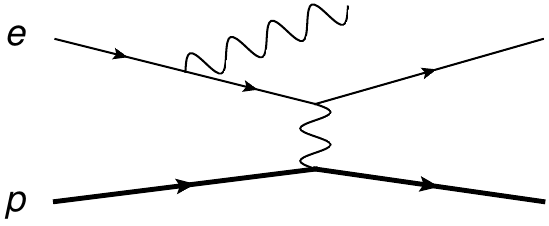,scale=1.0}  }
\end{minipage}
\begin{minipage}[t]{16.5 cm}
\caption{Initial state radiation in elastic electron-proton scattering.\label{fig:isr}}
\end{minipage}
\end{center}
\end{figure}

An experiment at MAMI is underway, aimed at measuring proton form-factors at very low momentum transfers by using a technique based on initial state radiation (ISR)~\cite{Mihovilovic2013}.   The idea is that ISR, Fig.~\ref{fig:isr}, degrades the energy of the incoming electron so that the momentum transfer $Q^2$ to the proton can be quite low.  The outgoing electron angle and energy are measured as usual, and with some modeling the diagram shown can be isolated from other radiative corrections and an accurate form factor obtained with $Q^2$ as low as $10^{-4}$ GeV$^2$.    There was a pilot measurement in 2010,  the full experiment has run in 2013, and the analysis is continuing.

\subsection{Electron deuteron scattering}


Electron-deuteron scattering data currently available does not by itself give a accurate enough determination of the deuteron charge radius to assess any discrepancy occasioned by the muonic deuterium Lamb shift.  Fortunately, the hydrogen-deuterium isotope shift allows using the proton radius measurement to obtain the deuteron radius to high accuracy~\cite{Parthey:2010aya}, as discussed in Sec.~\ref{sec:deuterium} above.  However, confirmation from direct scattering measurement is underway at Mainz~\cite{DistlerGriffioen}.  Data has been taken, and the analysis is ongoing.


\subsection{High precision Lamb shift in $e$-$H$}


This is a proposed measurement of the ordinary atomic hydrogen $n=2$ Lamb shift using the Ramsey method of separated oscillatory fields~\cite{VuthaHessels2012}. This new measurement,  with an anticipated uncertainty 5 times more accurate than existing measurements, along with existing precise atomic theory calculations, will by itself allow for a new determination of the proton charge radius to an accuracy of 0.6 percent.  Currently the quoted uncertainty in the proton charge radius from atomic spectroscopy comes from averaging over a large basket of measurements.  Here one will get a lower uncertainty result from just one measurement.


\subsection{New measurements of larger $e$-$H$ splittings at Garching}


Work is underway at the Max Plank Institute for Quantum Optics in Garching on a new precision measurement of the $2S$-$4P$ transition in atomic hydrogen.  This measurement, paired with the very accurate $1S$-$2S$ measurement can give very accurately both the Rydberg and the proton radius~\cite{Beyer:2013JPhCS.467a2003B,Parthey:2011lfa}.  Currently, as relevant also above, the individual electronic hydrogen spectroscopy measurements or pairs of measurements have proton radius uncertainties of more than 2\% and the small overall uncertainty comes from averaging over many measurements.  Here one will get and low uncertainty results from one measurement.


\subsection{New measurements of larger $e$-$H$ splittings at LKB, Paris}


Also in the category of new measurements of larger $e$-$H$ splitting is work underway on a precision measurement of the $1S$-$3S$ splitting in atomic hydrogen at the Laboratoire  Kastler Brossel (LKB) in Paris.  Again, this measurement by itself paired with the very accurate $1S$-$2S$ measurement will give very accurately both the Rydberg and the proton radius.  To get an idea of the precision necessary, the $1S$-$3S$ splitting is about $2.9 \times 10^{12}$ kHZ, and the difference between the predicted splittings with the CODATA radius and the $\mu$-$H$ radius is about $7.2$ kHz or a relative difference of about $2.5$ parts per trillion (ppt).  A $1\%$ accuracy for the proton radius will require a measurement to $1.8$ kHz ($0.6$ ppt) or less.  The 2010 result from the LKB for the same splitting, which will be superseded,  had a uncertainty limit of $13$ kHz~\cite{Arnoult:2010}.


\subsection{Alternative measurements of the Rydberg}


As discussed earlier, the Rydberg constant and the proton charge radius are determined in the same spectroscopy experiments.  If the muonic hydrogen Lamb shift  proton radius is used, the Rydberg constant would change by 4 standard deviations.  It may be possible to independently measure the Rydberg constant by using Rydberg states (very high-lying states) where the effect of the proton size is negligible.  Consideration of this possibility is underway at the National Insititute of Science and Technology (NIST, USA)~\cite{Guise:2012}.


\subsection{Trumuonium at JLab}


True muonium is the $\mu^+ \mu^-$ bound state.  If there is a exotic exchanged particle that couples to muons, it will affect the binding energy of true muonium.  The Heavy Photon Search experiment (HPS) at Hall B at Jefferson Lab will as a side project also have the possibility of finding true muonium~\cite{Phillips:truemuonium2012}.  There is hope of seeing 60-100 true muonium events;  it is of course some distance from just observing true muonium to measuring its spectrum accurately.

\section{Conclusion}
 
The proton charge radius puzzle remains a major problem.

The conflict is between measurements using electrons and the one muonic experiment, which measures the muonic hydrogen Lamb shift.  On the other hand, the muonic experiment has by far the smallest uncertainty limit of any proton radius measurement and while it is a difficult experiment to make work, it is does not seem to be a difficult experiment to analyze.  Further, the theoretical corrections have been reexamined and repeated by a number of worker since the announcement of the result, with no generally accepted problem has been found.

Perhaps the discrepancy is due to new physics.  Perhaps the explanation is an ordinary physics effect that has been missed.  Perhaps, the muonically measured radius will come to be the accepted number. Time will tell.

\medskip
\noindent \textit{Acknowledgments.}
I thank Jan Bernauer, Michael Distler, Richard Hill, Mikhail Gorchtein, Harald Griesshammar, Keith Griffioen, Miha Mihovilovic, Jerry Miller, Anuradha Misra, Asmita Mukherjee, Kostas Orginos, Vladimir Pascalutsa, Marc Vanderhaeghen, and Thomas Walcher for help and for useful conversations, and the National Science Foundation (USA) for support under grant PHY-1205905.

\bibliography{MuonicLamb2014}

\bibliographystyle{h-physrev5.bst}

\end{document}